\documentclass[superscriptaddress,amsfonts,amssymb,amsmath,twocolumn]{revtex4}
    \usepackage{epsfig}
    \usepackage{graphics}  
    \usepackage{amsmath}
    \usepackage{color}
    \usepackage{soul}
    \date{\today}
    
\begin{document}
    
    \title{Uniformly accelerated classical sources as limits of Unruh-DeWitt detectors}
    
    \author{Gabriel Cozzella}\email{gabriel.cozzella@unesp.br}    
    \affiliation{Instituto de F\'\i sica Te\'orica, 
    Universidade Estadual Paulista, Rua Dr.\ Bento Teobaldo Ferraz, 271, 01140-070, S\~ao Paulo, S\~ao Paulo, Brazil}
    
    \author{Stephen A. Fulling}\email{fulling@math.tamu.edu}    
    \affiliation{Department of Physics, Texas A\&M University, College Station, Texas, 77843-4242, USA}
    \affiliation{Department of Mathematics, Texas A\&M University, College Station, Texas, 77843-3368, USA}
    
    \author{Andr\'e G.\ S.\ Landulfo}\email{andre.landulfo@ufabc.edu.br}
    \affiliation{Centro de Ci\^encias Naturais e Humanas,
    Universidade Federal do ABC, Avenida dos Estados, 5001, 09210-580, Santo Andr\'e, S\~ao Paulo, Brazil}
    
    \author{George E.\ A.\ Matsas}\email{matsas@ift.unesp.br}
    \affiliation{Instituto de F\'\i sica Te\'orica, 
    Universidade Estadual Paulista, Rua Dr.\ Bento Teobaldo Ferraz, 271, 01140-070, S\~ao Paulo, S\~ao Paulo, Brazil}
    
        
    
    \begin{abstract} 
    Although the thermal and radiative effects associated with a two-level quantum system undergoing acceleration are now widely understood and accepted, a surprising amount of controversy still surrounds the simpler and older problem of an accelerated classical charge.  We argue that the analogy between these systems is more than superficial:  There is a sense
in which a ``UD detector'' in a quantized scalar field effectively acts as a classical source for that field if the splitting of its energy levels is so small as to be ignored.
 After showing explicitly that a detector with 
unresolved inner structure does behave as a structureless scalar 
source,  we use that analysis to rederive the scalar version of 
a previous analysis of the accelerated electromagnetic charge, without
appealing to the troublesome concept of ``zero-energy particles.'' 
Then we recover these results when the detector energy gap is taken to be zero 
from the beginning.
This vindicates the informal terminology 
 ``zero-frequency Rindler modes" as a shorthand
for ``Rindler modes with arbitrarily small energy."
In an appendix the mathematical behavior of the normal modes in the limit of small frequency 
is examined in more detail than before.
The vexed (and somewhat ambiguous) question of whether coaccelerating observers detect
the acceleration radiation can then be studied on a sound basis.
    \end{abstract}
    
    \maketitle
    
    \section{Introduction} \label{sec:I} 
    
    In 1992 it was shown~\cite{HMS92-2} that the ordinary  emission of a photon from a uniformly accelerated classical charge in the Minkowski vacuum corresponds to either the absorption from or the emission to the Unruh thermal bath of a zero-energy Rindler photon (as defined by uniformly accelerated observers). This fact is strikingly parallel to the situation for Unruh-DeWitt detectors~\cite{UW84}.
    
     We recall that since the Rindler frequency $\omega_R $ and transverse momentum ${\bf k}_\bot $ are not constrained by any dispersion relation, zero-energy Rindler photons exist for arbitrarily large  ${\bf k}_\bot$ (whereas zero-energy Minkowski photons are rather trivial)~\cite{polar}. 
    Nevertheless, it can be argued that 
      uniformly accelerated (Rindler) observers do not experience these photons as ``radiation''\negthinspace, for two reasons:  First, they represent fields that are static (with respect to Rindler time).  Second, 
        zero-energy Rindler photon modes concentrate near the horizon, so that localized Rindler observers with finite proper acceleration  barely have an opportunity to interact with them.
      These observations are  in harmony with classical-electrodynamics results according to which uniformly accelerated charges radiate for inertial observers but do not for coaccelerating ones~\cite{FR60,B80}. The scalar analogs of the main conclusions of Refs.~\cite{HMS92-2} and~\cite{B80} were worked out in Ref.~\cite{RW94}. Furthermore, the conclusions reached in Ref.~\cite{HMS92-2} were strengthened by a recent nonperturbative 
      calculation~\cite{LFM19}, which derived the usual classical Larmor radiation, 
      emitted by a uniformly accelerated scalar source, exclusively from the zero-energy  Rindler modes. 
    
    Despite this, Ref.~\cite{HMS92-2} still provokes debate. One reason, which is not our main concern in this paper,  is that the calculations in the Rindler frame made use of an 
   extended  oscillating dipole regularization, a simple oscillating charge being inconsistent
    with the charge conservation required by the electromagnetic theory.
    This issue does not arise in the scalar analog.
    Furthermore, either adding some transverse velocity to the charge (which makes it to interact also with nonzero energy modes) \cite{CLMV17,CLMV18} or adding a small Proca mass to the field~\cite{RF} in intermediate stages of the calculation 
    makes it possible to study an oscillating charge strength without introducing a dipole.
    
    
    
   
   A second stumbling block has been the unfamiliarity of the notion of zero-energy
   (or zero-frequency) modes.  To address it, we present here a different approach.
    As in Refs.~\cite{RW94,LFM19} we replace the electric charge by a pointlike scalar source in order to simplify the technical analysis without sacrificing the conceptual discussion.
     More importantly,  in order to circumvent the introduction of any explicit regularization, we endow the scalar source with an internal-energy degree of freedom. The energy gap is assumed to be nonresolvable by any available technology. 
  Composite particles behave as elementary ones insofar as one does not
observe their inner structure; for example, at a certain mesoscopic level a hydrogen atom can be treated as a point particle even though its ground state has hyperfine structure (with a very small energy splitting).  
In the same way, our two-level scalar
system will behave as a structureless source insofar as one does not have enough precision to probe the internal energy gap.  
It interacts, however, with field modes of small but nonvanishing energy.
    
    We emphasize that all these different approaches to the idealized limit of an eternally accelerating point charge yield essentially identical results in the end, thereby vindicating each other.
    
    In Sec.~\ref{sec:II} we present the physical setup, calculate the emission rate of a scalar source with unresolved inner structure, and  show how  inertial and Rindler observers' results are harmonized. 
    We confirm that a two-level scalar system with unresolved energy gap behaves as a structureless source.
     In Sec.~\ref{sec:new} and Appendix \ref{app:gapless} we 
     investigate what happens when the splitting of the UD system's energy levels is 
     exactly zero. 
    Our conclusions are summarized in Sec.~\ref{sec:IV}. 
    We adopt  $(+,-,-,-)$ for the metric signature, and natural units, $\hbar=c=k_{\rm B}=1$, unless stated otherwise. 
     \section{Emission rate from Unruh--DeWitt detectors with unobserved internal structure in the Rindler frame}\label{sec:II} 

\subsection{Detector model}\label{ssec:IIA}
    
    For the sake of simplicity, we replace the electric charge by a pointlike scalar source as in~\cite{RW94,LFM19}. Instead of a structureless source, we  consider a two-level scalar system --- also known as Unruh--DeWitt (UD) detector~\cite{U76,D78}. From a mathematical perspective this will turn out to be  convenient, while from a physical perspective, it will not affect the results as far as we assume that the energy gap is unresolved by any technology. Indeed, as  will be shown later, an Unruh--DeWitt detector with inner energy levels $E$ and $E + \Delta E$ will radiate as a structureless source (with mass $m=E/c^2$) provided the energy gap $\Delta E$ cannot be resolved. 
    
    A pointlike UD detector is represented by a monopole operator ${\hat m} (\tau )$. It  acts on the orthonormal detector energy eigenstates $| E_\pm \rangle$ as 
    \begin{equation}
    \hat m(0) |E_\pm \rangle = |E_\mp \rangle ,
    \label{monopole}
    \end{equation}
    where $E_+>E_-$,
    $\langle E_\pm | E_\mp\rangle =0$, $\langle E_\pm | E_\pm\rangle =1$.
    We let $\Delta E\equiv E_+ - E_-$. 
    
    The detector will be minimally coupled to a massless scalar field $\hat \phi$ through the interaction action 
    \begin{equation}
    {\cal S}_I=\int d\tau c(\tau) \hat m(\tau) \hat \phi [x^\mu (\tau)], 
    \label{action}
    \end{equation}
    where $x^\mu (\tau) $ is the detector worldline,  $\tau$ is its proper time, and $\hat{\phi}$ satisfies
    \begin{equation}
        \Box \hat \phi =0.
        \label{KG}
    \end{equation}
    The switching function $c(\tau)$ is assumed  (i)~to be at least continuous to avoid the appearance of divergences (arising because it is physically impossible to instantaneously switch on/off a detector~\cite{HMP93,S04,LS08}) and (ii)~to have compact support to keep the detector switched on  for only a finite amount of time. 
    \begin{figure}
       \centering
       \includegraphics[width=60mm]{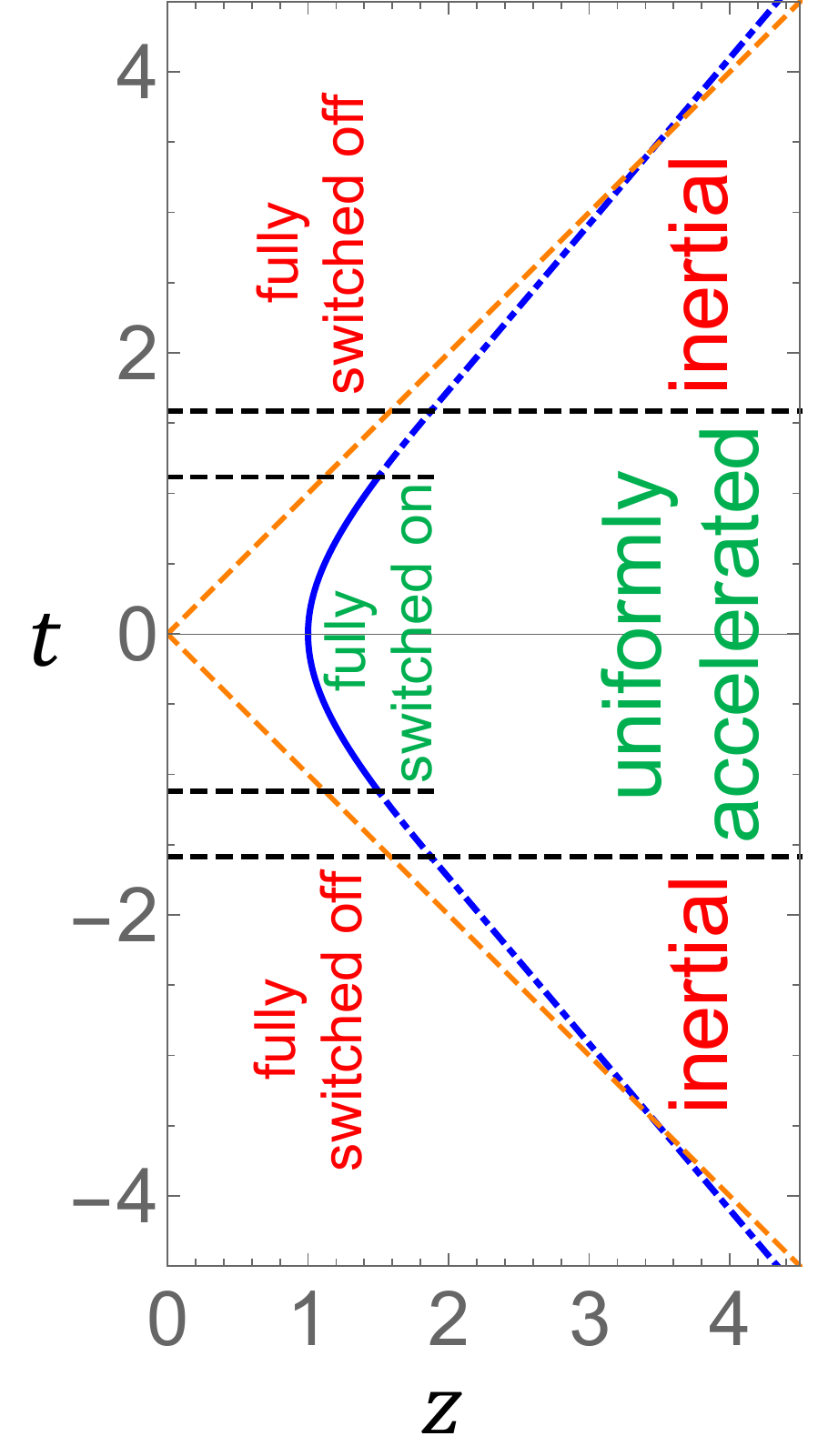}
        \caption{The UD detector enters the  Rindler wedge (the region to the right of the dashed diagonal lines) inertially when it is still switched off (dot-dashed line). At some point, the detector is uniformly accelerated by an external agent. After this, the detector is switched on continuously. The detector stays uniformly accelerated and fully switched on for some proper time $T$ (full line). After this, the whole process is reversed: first, the detector  is switched off, and, next, it is made inertial before leaving the wedge. }
       \label{Fig1}
    \end{figure}
    
    The physical picture is depicted in Fig.~\ref{Fig1}: the UD detector enters the Rindler wedge switched off and free. Then, some external agent uniformly accelerates it, after which the detector is switched on. A real parameter $\alpha$ regulates how fast the detector is switched on/off: the larger the $\alpha$ the faster the detector is switched on/off ({\em i.e.,} the smaller the switching time). The detector stays uniformly accelerated and fully switched on for some proper time $T$. The whole process is invariant under time reflection. 
    
    This setup  aims to disentangle our analysis from the interesting (but academic) question of whether ({\em eternally}) uniformly accelerated sources radiate with respect to inertial observers (see, {\em e.g.,} p.~391 of Ref.~\cite{SDMT98}). We assume that there is no dispute about the fact that accelerated electric charges (and corresponding scalar sources) with physical trajectories do radiate according to inertial observers. Indeed, this is used as a benchmark in the discussion below.

    In order to calculate the radiation emitted by the detector with respect to Rindler observers, we must first quantize the scalar field according to them. With no loss of generality, we  choose the detector to be in the right Rindler wedge of the Minkowski spacetime, $z >|t| $, as shown in Fig.~\ref{Fig1}. (From here on, by ``Rindler wedge" we mean ``right Rindler wedge" unless stated otherwise.)  It is convenient to use Rindler coordinates $(\tau, \xi, x, y)$ to cover the wedge. They are related to the usual Cartesian coordinates $(t, x, y, z)$ by
    \begin{equation}
    t = ({e^{a\xi}}/{a}) \sinh a\tau, \;
    z = ({e^{a\xi}}/{a}) \cosh a\tau.
    \label{RC}
    \end{equation}
    Here, $a > 0$ and $-\infty < \tau,\xi , x, y <\infty $. 
    The metric of Minkowski space, restricted to the Rindler wedge,  takes in these coordinates the form
    \begin{equation}
    ds^2 = e^{2a\xi} (d\tau^2 - d\xi^2) -dx^2 -dy^2 .
    \label{RLE}
    \end{equation}
    The worldlines of Rindler observers with constant proper acceleration $ a e^{-a \xi}$ are given by $\xi, x, y = {\rm const}$ and  consist of a congruence covering the whole Rindler wedge ---
    the orbits of a timelike Killing vector. We recall that the Rindler wedge is a globally hyperbolic and static
    spacetime in its own right, and so Rindler observers are perfectly eligible to analyze the radiation emitted from UD detectors which are made active inside the wedge.  
    
    The  scalar field $\hat \phi$ can be Fourier-decomposed in the Rindler wedge in terms of a complete set of (asymptotically well behaved) orthonormal modes $\{u_{\omega_R {\bf k}_\bot} ,u_{\omega_R {\bf k}_\bot}^*\}$ satisfying Eq.~(\ref{KG}) as (see, {\em e.g.,} Ref.~\cite{CHM08} for a review, and references therein)
    \begin{equation}
    \hat \phi (x^\nu ) = \int d^2{\bf k}_\bot
    \int_{0}^{+\infty} d\omega_R
    \left\{{\hat a^R}_{ \omega_R {\bf k}_\bot}
           u_{\omega_R {\bf k}_\bot }
           (x^\nu ) + {\rm H.c.} \right\},
    \label{REF}
    \end{equation}
    where 
    ${\hat a^R}_{ \omega_R {\bf k}_\bot}$
    and
    $\hat a^{R \dagger}_{ \omega_R {\bf k}_\bot}$ 
    are the annihilation and creation operators of Rindler modes, 
    respectively, 
    ${\bf k}_\bot \equiv (k_x,k_y)$, $k_\bot = |{\bf k}_\bot|$, and
    \begin{eqnarray}
    u_{\omega_R {\bf k}_\bot} 
    &=& 
    \left[ \frac{\sinh ({\pi \omega_R}/{a})}
    {4\pi^4 a}\right]^{ \frac{1}{2}} K_{i\omega_R/a}
    \left(\frac{k_\bot}{a} e^{a\xi } \right)
    \nonumber \\
    &\times&
    e^{i {\bf k}_\bot\cdot {\bf x}_\bot - i \omega_R \tau} .
    \label{RM}
    \end{eqnarray}
    
      \subsection{Analysis in terms of Rindler particles}\label{ssec:IIB}
    
    Now, let us calculate the emission rate from the UD detector assuming that the field 
    starts in the Minkowski vacuum. Since we will be interested in the radiation emitted by an UD detector with some unresolved inner structure, we will assume it to be initially in the
     mixed state given by the density matrix
    \begin{equation}
       \hat\rho = A|E_+ \rangle\langle E_+| + B|E_- \rangle\langle E_-|, \quad  A + B = 1. 
    \label{rho}
    \end{equation}
  We note that for $A=B=\frac12$ we have the maximum entropy (ignorance) state.
    Also, our final result (\ref{totalMinkowskirate}) and the discussion that follows it
    would not be affected had we chosen $\hat\rho$ to include interference terms.
    (See Eq.~(\ref{qubitstatepast}), where $\hat\rho$ is chosen to  represent a
     general state.)
    Thus, we must calculate the emission probability of a Rindler particle accompanied by  either deexcitation or excitation of the detector, and average, in the end, over the detector's initial states. (We recall that,  according to the action~(\ref{action}), a particle emission must be always accompanied  by either excitation or deexcitation of the detector.) 
    
    In first-order perturbation theory,
    the  emission probability of a Rindler particle with  simultaneous deexcitation (deexc) or excitation (exc)  is  
    \begin{equation}
    {\cal P}^{\rm (de)exc}_{\rm em} 
    = \int dW^{\rm (de)exc}_{\rm em} \left( 1 + \frac {1}{e^{\omega_R /T_U} - 1} \right),
    \label{PMEM2}
    \end{equation}
    where we recall that according to Rindler observers the Minkowski vacuum corresponds to a thermal state at the Unruh temperature~\cite{U76}
    \begin{equation}
    T_U=a/(2\pi).
    \label{T_U}
    \end{equation}
    The first and second terms inside the parentheses in Eq.~(\ref{PMEM2}) correspond to spontaneous and induced emissions, respectively. Here
    \begin{equation}
    dW^{\rm (de)exc}_{\rm em} = \vert {\cal A}_{\rm em}^{\rm (de)exc} \vert ^2 d^2{\bf k}_\bot d\omega_R
    \label{DWREM2}
    \end {equation}
    are the vacuum differential emission probabilities, where
    \begin{eqnarray}
    {\cal A}_{\rm em}^{\rm deexc} &&= \langle \omega_R {\bf k}_\bot \vert \otimes \langle E_-
                 \vert \; {\cal S}_I\; \vert
                 E_+\rangle  \otimes \vert  0_R \rangle,
    \label{AREM2}
    \\
    {\cal A}_{\rm em}^{\rm exc} &&= \langle \omega_R {\bf k}_\bot \vert  \otimes \langle E_+
                 \vert \; {\cal S}_I\; \vert
                 E_-\rangle  \otimes \vert  0_R \rangle.
    \label{AREM0}
    \end{eqnarray}
    The total emission probability is obtained by averaging on the detector initial states:
    \begin{equation}
     {\cal P}^{\rm total}_{\rm em} = A {\cal P}^{\rm deexc}_{\rm em} + B {\cal P}^{\rm exc}_{\rm em}, 
     \quad A+B=1.   
        \label{emission}
    \end{equation}
    
    Because we are interested ultimately in whether a uniformly accelerated source radiates with respect to coaccelerating observers, we will focus on the regime where $T$ is larger than any other scale of the problem: 
    \begin{equation}
        T \gg \alpha^{-1} ,a^{-1} ,\Delta E^{-1}.
        \label{regime}
    \end{equation} 
    In this regime, it is convenient to schematically cast ${\cal P}^{\rm deexc}_{\rm em}$ and ${\cal P}^{\rm exc}_{\rm em}$ as 
    \begin{equation}
    {\cal P}^{\rm deexc}_{\rm em} \approx 
    {\cal P}^{{\rm deexc}(0)}_{\rm em} \left( \frac{\Delta E}{a}, aT \right)
    +
    {\cal P}^{{\rm deexc}(1)}_{\rm em} \left(\frac{\Delta E}{a}, \frac{\alpha}{a} \right)
        \label{PEMDEEXC} 
    \end{equation}
    and
    \begin{equation}
    {\cal P}^{\rm exc}_{\rm em} \approx 
    {\cal P}^{{\rm exc}(1)}_{\rm em} \left( \frac{\Delta E}{a}, \frac{\alpha}{a} \right),
        \label{PEMEXC} 
    \end{equation}
    respectively. Equations~(\ref{PEMDEEXC}) and~(\ref{PEMEXC}) can be understood  under physical grounds (or explicitly derived~\cite{HMP93} assuming some switching function) as follows. Firstly, we note that since the detector is at rest with respect to Rindler observers (whenever it is switched on), energy conservation enforces that any nonvanishing contribution to the emission probability is  due either to the detector deexcitation or  to the switching process realized by an external agent.  Thus,  Eq.~(\ref{PEMEXC}) 
    holds because any contribution to the emission probability accompanied by detector excitation must come from the switching process.  Equation~(\ref{PEMDEEXC}),
    in turn,  expresses the fact that for large enough $T$, ${\cal P}^{{\rm deexc}}_{\rm em}$ can be decomposed into two independent contributions: (i)~a dominant one, ${\cal P}^{{\rm deexc}(0)}_{\rm em}$, just associated with the long acceleration period $T$ --- the larger the $T$ the larger the ${\cal P}^{{\rm deexc}(0)}_{\rm em}$ --- and (ii)~a minor one, ${\cal P}^{{\rm deexc}(1)}_{\rm em}$, ruled by $\alpha$, which  is a transient response caused by the switching process. Both these contributions must depend on $\Delta E$, which carries information on the detector itself, while the acceleration $a$ is properly combined with $T$, $\alpha$, and $\Delta E$ to comply with the fact that ${\cal P}^{{\rm deexc}}_{\rm em}$ is dimensionless.
    
    As a result,  Eq.~(\ref{emission}) in the regime~(\ref{regime})  is approximated  as
    \begin{equation}
     {\cal P}^{\rm total}_{\rm em} \approx A\, {\cal P}^{{\rm deexc}(0)}_{\rm em}\left( \frac{\Delta E}{a}, aT \right).   
        \label{emission2}
    \end{equation}
    The fact that ${\cal P}^{{\rm deexc}(0)}_{\rm em}$ does not contain information on $\alpha$ is particularly handy, since Eq.~(\ref{emission2}) can be calculated as if the detector were permanently switched on. It is straightforward, hence, to obtain the corresponding  amplitude from Eq.~(\ref{AREM2}) (up to an arbitrary global phase): 
    \begin{eqnarray}
    {\cal A}^{{\rm deexc}(0)}_{\rm em} 
    &=& c_0 
    \left[ 
    \frac{\sinh (\pi \omega_R/a)}{\pi^2 a } 
    \right]^{1/2}  K_{i\omega_R /a}
    \left(\frac{k_\bot}{a}  \right)
    \nonumber \\
    &\times& 
    \delta(\omega_R - \Delta E),
    \label{AREM4}
    \end{eqnarray}
    where $c_0 = {\rm const}$ corresponds to the value of $c(\tau)$ as the detector is fully switched on, and we have used that the monopole operator evolves in time as
    \begin{equation}
    \hat m (\tau) = e^{i{\hat H}_0\tau} \hat m(0) e^{-i {\hat H}_0\tau}.
    \label{M}
    \end{equation}
    Here, ${\hat H}_0$ is the detector free Hamiltonian: 
    $$
    {\hat H}_0\vert E_\pm \rangle = E_\pm \vert  E_\pm \rangle
    $$
     and for the sake of simplicity we have chosen $|\langle E_- | \hat m_0 |  E_+ \rangle | \equiv 1$.  The delta function $\delta(\omega_R - \Delta E)$ in Eq.~(\ref{AREM4}) expresses the fact that for large enough $T$, the total emission probability, ${\cal P}^{\rm total}_{\rm em}$, is quite dominated by the emission of Rindler particles with $\omega_R=\Delta E$. 
    
    It is important to note, now, that for detectors with $\Delta E \ll a$, the emitted particles ($\omega_R = \Delta E \ll a$) concentrate near the horizon.
    This tendency is visible in  Fig.~\ref{fig2}, which compares modes for two different frequencies $\omega_R$;  it is examined in more detail in Appendix \ref{app:zeromode}.
     It should 
    come as no surprise, then,  that the smaller  $\Delta E$ is, the more difficult it is for the detector (with fixed finite $a$) to radiate into the vacuum. 
    And indeed, ${\Gamma}^{\rm tot}_{\rm em}$ would approach zero in that limit if the
    physically relevant field state were the ``Rindler vacuum'' 
    (corresponding to omission
    of the Planckian term in Eq.~(\ref{PMABS0})).
        The fact is, however, that the Rindler observers are immersed in the Unruh thermal bath,
        and that changes dramatically this conclusion. In order to see this, let us explicitly calculate the dominant emission contribution,
    \begin{equation}
    {\cal P}^{{\rm deexc} (0)}_{\rm em} 
    =
    \int dW^{{\rm deexc}(0)}_{\rm em} \left( 1 + \frac {1}{e^{\omega_R /T_U} - 1} \right),
    \label{PMABS0}
    \end{equation}
    where $dW^{{\rm deexc} (0)} \equiv |{\cal A}^{{\rm deexc}(0)}_{\rm em} |^2 d^2{\bf k}_\bot d\omega_R $. 
    By using 
    \begin{equation}
    \int dk_\bot^2 K^2_ {i\omega/a} (k_\bot/a) = \frac{\pi^2 a \omega}{\sinh (\pi \omega/a)}
    \end{equation}
    and
    \begin{equation}
    T = \lim_{\omega_R \to 0} \int_{-\infty}^{\infty} d\tau e^{-i\omega_R \tau} = 2 \pi \delta(0), 
    \label{T}
    \end{equation}
    we have
    \begin{eqnarray}
    {\cal P}^{{\rm deexc} (0)}_{\rm em} 
    &=&
    T c_0^2
    \int_0^{\infty} d\omega_R \frac{ \omega_R}{2 \pi}  
    \left( 1+ \frac {1}{e^{2\pi \omega_R /a} - 1} \right)
    \nonumber \\
    &\times& \delta (\omega_R-\Delta E)
    \nonumber \\
    &=& 
    c_0^2 T \frac{\Delta E}{2 \pi} \left( 1 + \frac {1}{e^{2\pi \Delta E/a} -1} \right).
    \label{PMABS}
    \end{eqnarray}
    Now, the assumption that the inner structure of the detector is nonobservable begs for the $\Delta E/a \to 0$ limit in Eq.~(\ref{PMABS}).
     The Planckian term's contribution survives in the limit, and one obtains
    \begin{equation}
    {\cal P}^{{\rm deexc} (0)}_{\rm em} 
    =
    T \frac{c_0^2 a}{4 \pi^2}.
    \label{Gamma^L3}
    \end{equation}
    Next, we use Eq.~(\ref{emission2}) to obtain the total emission rate of Rindler particles: 
    \begin{equation}
    \Gamma^{\rm total}_{\rm em} 
    \equiv 
    \frac{{\cal P}^{\rm total}_{\rm em}}{T} 
    = 
    A \frac{c_0^2 a}{4 \pi^2},
        \label{totalemissionrate0}
    \end{equation}
    where ``$\approx$" in Eq.~(\ref{emission2}) was replaced by ``$=$", as we assume $T$ to be arbitrarily large.
    \begin{figure}
       \centering
       \includegraphics[width=85mm]{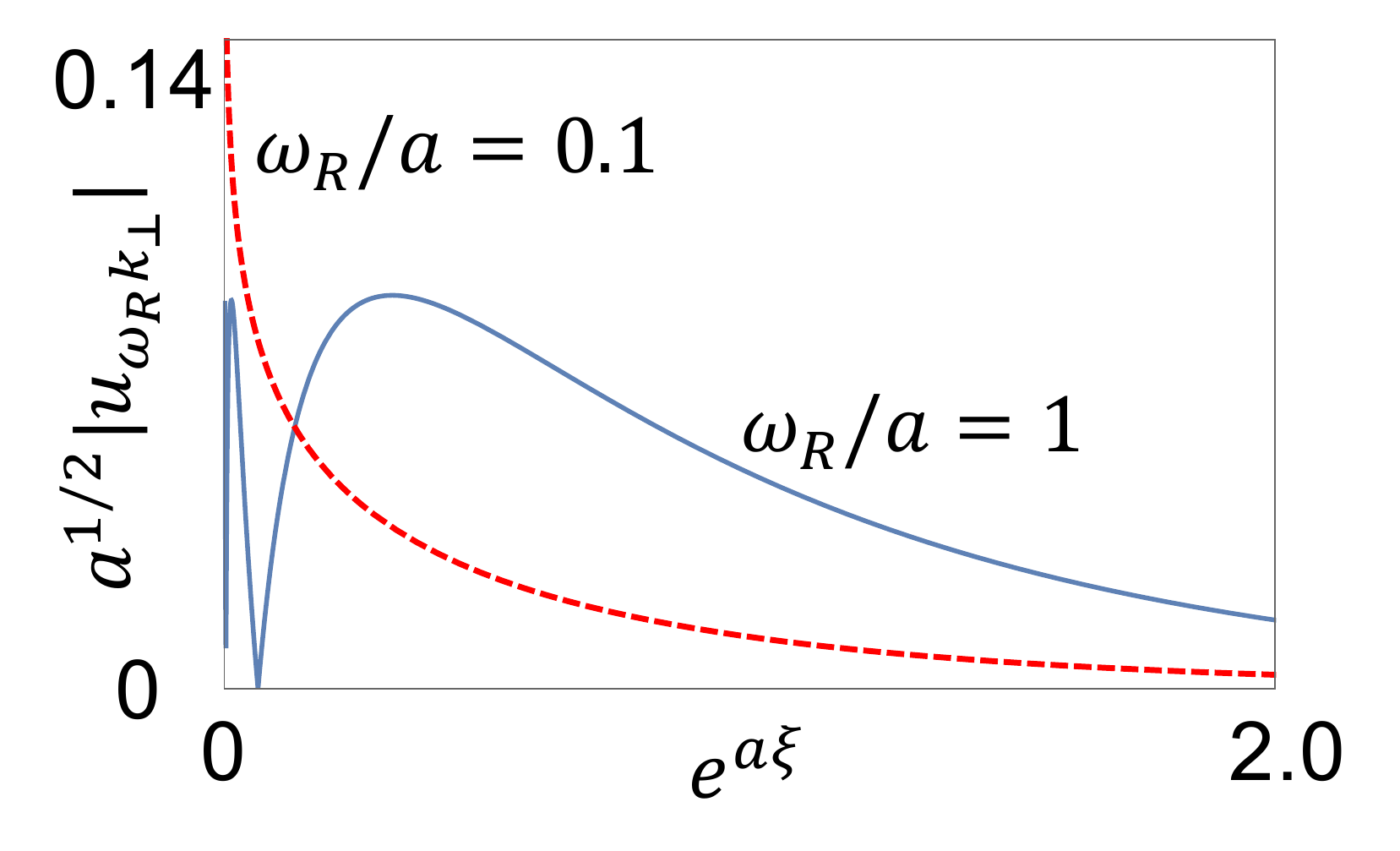}
        \caption{ The plot illustrates that the smaller the $\omega_R/a$ the more the Rindler mode concentrates on the horizon, $\xi\to -\infty$. The curves in the plot assume $a=k_\bot=1$.}
       \label{fig2}
    \end{figure}
    
    Thus the Unruh thermal-bath factor is essential for obtaining the nonvanishing
    $\Gamma^{\rm total}_{\rm em} $, which is  the correct result, needed for agreement
    with Larmor's classical bremsstrahlung formula and rederived in the next section
    by an independent method. 
     Nevertheless, as explained above, the corresponding Rindler particles emitted (with arbitrarily small $\omega_R$) concentrate near the horizon and, thus, are practically inaccessible to physical Rindler observers. 
     
     Finally, because Eq.~(\ref{totalemissionrate0}) does not carry any information about $\Delta E$, we shall identify it with the emission rate of a structureless scalar source, as well. This will be explicitly shown to be correct further on.
    
    \subsection{Analysis in terms of Minkowski particles} \label{ssec:IIC}

    In order to harmonize our previous conclusion with the emission rate of  ``Minkowski''
     particles (those defined by inertial observers), we must calculate the corresponding absorption rates of Rindler particles. 
    
    The absorption probability of a Rindler particle from the Unruh thermal bath with  simultaneous detector deexcitation or excitation is 
    \begin{equation}
    {\cal P}^{\rm (de)exc}_{\rm abs} 
    = \int dW^{\rm (de)exc}_{\rm abs} \frac {1}{e^{\omega_R /T_U} - 1},
    \label{PMABS2}
    \end{equation}
    where
    \begin{equation}
    dW^{\rm (de)exc}_{\rm abs} = \vert {\cal A}_{\rm abs}^{\rm (de)exc} \vert ^2 d^2{\bf k}_\bot d\omega_R
    \label{DWRABS2}
    \end {equation}
    and
    \begin{eqnarray}
    {\cal A}_{\rm abs}^{\rm deexc} &&= \langle  0_R \vert \otimes \langle E_-
                 \vert \; {\cal S}_I\; \vert
                 E_+\rangle  \otimes \vert \omega_R {\bf k}_\bot  \rangle,
    \label{AREM}
    \\
    {\cal A}_{\rm abs}^{\rm exc} &&= \langle 0_R  \vert  \otimes \langle E_+
                 \vert \; {\cal S}_I\; \vert
                 E_-\rangle  \otimes \vert  \omega_R {\bf k}_\bot  \rangle.
    \label{ARABS}
    \end{eqnarray}
    The transition amplitudes above are obtained from the previous section's results by recalling that
    \begin{equation}
    |{\cal A}_{\rm abs}^{\rm deexc}| = |{\cal A}_{\rm em}^{\rm exc}|, \quad
    |{\cal A}_{\rm abs}^{\rm exc}| = |{\cal A}_{\rm em}^{\rm deexc}|,
   \label{4rates} \end{equation}
    so that
    \begin{equation}
    dW^{\rm deexc}_{\rm abs} = dW^{\rm exc}_{\rm em}, \quad  
    dW^{\rm exc}_{\rm abs} = dW^{\rm deexc}_{\rm em}.
    \label{DWABS3}
    \end {equation}
    It is straightforward, now, to calculate the  total absorption probability
    \begin{equation}
     {\cal P}^{\rm total}_{\rm abs} = A {\cal P}^{\rm deexc}_{\rm abs} + B {\cal P}^{\rm exc}_{\rm abs}, \quad A+B=1,   
        \label{absorption}
    \end{equation}
    using Eq.~(\ref{PMABS2}). Indeed, the corresponding total absorption rate of Rindler particles in the regime~(\ref{regime}) turns out to be given precisely by Eq.~(\ref{totalemissionrate0}) with $A$ replaced by   $B$:
    \begin{equation}
    \Gamma^{\rm total}_{\rm abs}
    \equiv 
    \frac{{\cal P}^{\rm total}_{\rm abs}}{T} 
    = B \frac{c_0^2 a}{4 \pi^2}.
        \label{totalabsorptionrate0}
    \end{equation}
    
    Now, because each {\em absorption} and {\em emission} of a Rindler 
particle uniquely corresponds to the {\em emission} of a Minkowski 
particle~\cite{UW84}, we can infer that the total emission rate of 
usual Minkowski particles from our UD detector with unresolved inner 
structures is   
    \begin{equation}
    \Gamma^{\rm total} 
    \equiv 
    \Gamma^{\rm total}_{\rm abs} + \Gamma^{\rm total}_{\rm em} 
    = \frac{c_0^2 a}{4 \pi^2}.
        \label{totalMinkowskirate}
    \end{equation}
    Again, because Eq.~(\ref{totalMinkowskirate}) does not carry any 
information about $\Delta E$, we identify it with the emission rate of 
Minkowski particles from a uniformly accelerated {\em 
structureless\/} scalar source. It happens that a straightforward 
inertial-frame calculation confirms it~\cite{RW94}. 
    
    Thus, a uniformly accelerated UD detector with unresolved inner 
structure behaves as a structureless scalar source, indeed. In 
 Appendix \ref{app:gapless} we offer an independent 
nonperturbative  calculation, 
along  the same  lines  of Ref.~\cite{LFM19}, where the detector is assumed to have $\Delta E= 0$ from the beginning, leading to the same result as Eq.~(\ref{totalMinkowskirate}). The calculation also makes  manifest once more that only zero-frequency Rindler modes contribute to the corresponding Larmor radiation.
    
    Reconciling the Minkowski picture of bremsstrahlung with the null result for
    radiation into the Rindler vacuum is an interesting problem, but it cannot be addressed here.
    It is meaningless until the Rindler vacuum state has somehow been extended into
    a state on the entire Minkowski space-time.  It is not obvious that such an extension
    exists, and it would surely be nonunique, as well as singular on the 
    horizon~\cite{P93}.

     \section{Gapless ``detectors''} \label{sec:new}
     
     In the two previous sections we have assumed that $\Delta E$, although positive, is so small that the
     internal structure and state of the detector are in practice unobservable.  One might question
     whether the system deserves to be called a ``detector'' under these conditions.
     Nevertheless, its effects on the quantized field are objective and observable.

     Another approach is to  take the limit $\Delta E \to 0$ in the  model 
     itself rather than the solution.  In that limit, $E_+ = E_-$, so the notation in and below
     Eq.\ (\ref{monopole}) becomes inappropriate.  Instead, we denote two basis states
     by $|+\rangle$ and $|-\rangle$:
       \begin{equation}
    \hat m(0) |\pm\rangle = |\mp \rangle , \quad
      \langle \pm | \mp\rangle =0,\quad \langle \pm | \pm\rangle =1.
    \label{monopole0}
    \end{equation}
    
    What happens when the calculations of Sec.\ \ref{sec:II} 
    are repeated in this context?
    Energy conservation can no longer be used to rule out simultaneous excitation and emission, or simultaneous deexcitation and absorption; both of the terms in Eq.\ (\ref{emission}),
   and all four in Eq.\ (\ref{4rates}), are nontrivial.  
    On the other hand, because only positive values  of $\omega_R$ contribute in
    Eq.\ (\ref{PMABS}) and similar integrals, the delta function $\delta(\omega_R)$
    is effectively multiplied by $\frac12$.
  Therefore, the results (\ref{totalemissionrate0}), (\ref{totalabsorptionrate0}),   (\ref{totalMinkowskirate}) of the calculation are the same as before.
  
     A further step of simplification is to change the interaction so that the detector's
     internal state does not change at all:
        \begin{equation}
    \hat m(0) |\pm\rangle = |\pm \rangle .
     \label{electronwithspin}
    \end{equation}
    The effect on the field is the same as before. 
    
    Finally, since the detector's 2-dimensional space of ground states has become superfluous, we can make it 1-dimensional:
    \begin{equation}
    \hat m(0) |E\rangle = |E \rangle.
\label{scalarelectron}
\end{equation}
Again perturbation theory predicts exactly the same effect of an accelerated ``detector''
on the field.  In all four stages of degeneration of the detector, the result is the same, (\ref{totalMinkowskirate}):
   \begin{equation}
 \Gamma^{\rm total}     = \frac{c_0^2 a}{4 \pi^2}
        \label{totalrate}
    \end{equation}
(with the same numerical coefficient).
In the final stage, however, the UD system has been replaced by a scalar source;
the calculations are just another variant of those in Refs.\ \cite{RW94} and \cite{LFM19}.

Thinking of system (\ref{scalarelectron}) as a scalar version of an electron, and then 
stepping back one stage, one sees that system (\ref{electronwithspin}) is merely
an electron with spin.  In the classical theory of bremsstrahlung, the spin plays no role.
(In particular, the electron's internal spin degree of freedom does not multiply
the emission rate by 2.)  In statistical mechanics, however, the existence of spin
modifies the entropy of a large ensemble, or gas, of electrons.
System (\ref{scalarelectron}) is indisputably ``structureless'';
in cases (\ref{electronwithspin}) and (\ref{monopole0}) the system is effectively
structureless, as the structure is irrelevant.

    \section{Conclusions}\label{sec:IV} 
    
    We have used UD detectors with an unresolved inner structure to illuminate the question of whether and how uniformly accelerated scalar sources 
    (and hence, by analogy, electrical charges)
    radiate to coaccelerating observers. We have shown that uniformly accelerated detectors with some unresolved inner structure emit only  Rindler particles with arbitrarily small frequency. It is concluded, then, that uniformly accelerated pointlike structureless sources emit only zero-energy Rindler particles. This is in harmony with classical-electrodynamics literature according to which uniformly accelerated electric charges do not radiate to coaccelerating observers~\cite{FR60,B80}. This is so partly because zero-frequency fields are static (hence not radiative in the usual sense), but also because zero-frequency Rindler modes 
    concentrate near the horizon, where physical Rindler observers ({\em i.e.,} Rindler observers with finite proper acceleration) have difficulty probing. (This 
    suppression of modes of minuscule energy is even more pronounced for quanta of a massive field~\cite{CCMV02,KSAD18}.)
    
    Our Rindler-observer results are also in harmony with independent calculations for the corresponding Minkowski-emission rate (as defined by inertial observers)~\cite{RW94}, since this equals the combined emission and absorption rates of Rindler particles (as it should). This is analogous to the result obtained in the electromagnetic case~\cite{HMS92-2}: {\em The emission of an ordinary photon from a uniformly accelerated classical electric charge in the Minkowski vacuum corresponds to either  absorption or emission of a zero-energy Rindler photon}.   
    
    We hope that the present paper helps to clarify the long-standing issue about whether uniformly accelerated sources radiate for coaccelerating observers, and strengthens the conclusion reached in Refs.~\cite{CLMV17,CLMV18} that {\em under proper conditions}~\cite{note} the observation of Larmor radiation  is a piece of circumstantial evidence in favor of the Unruh thermal bath. The link between the {\em classical\/} Larmor radiation (cast in terms of {\em classical\/} energy flux) and the {\em quantum\/} Unruh effect (associated with a thermal state of {\em quantum particles\/}) is provided by the simple Planck--Einstein relation $E= h \nu$. Classical phenomena must be described in terms of frequency, 
    not particle energy, and wave intensity or flux, not particle number. 
     For  more general discussions on the classical aspects of the Unruh effect see Refs.~\cite{HM93,PV99}. 
     Larmor radiation can be seen as a shadow of the Unruh thermal bath in Plato's cave: it reveals something about the true thing but not the whole thing (as always). 
    
    
    \acknowledgments
    We  thank Don Page, Bill Unruh, Paul Davies, Benito Juarez-Aubry, 
Eduardo Mart\'\i n-Martinez, and Jos\'e Ram\'on for discussions. G.~M. is deeply indebted to Atsushi Higuchi and Daniel Sudarsky for illuminating discussions on this and related questions over the years. G.~C. and G.~M. were fully and partially supported by S\~ao Paulo Research Foundation (FAPESP) under Grant 2016/08025-0 and Conselho Nacional de Desenvolvimento Cient\'\i fico e Tecnol\'ogico (CNPq) under Grant 301544/2018-2, respectively.
    
 \appendix
\section{Radiation emitted by a uniformly accelerated gapless UD detector}
\label{app:gapless}

Here we show explicitly by means of an exact (nonperturbative) calculation that a gapless UD detector with structure (\ref{monopole0}) 
emits like a classical {\em structureless\/} scalar source. 

In what follows, $\mathfrak{t}$ is a real parameter which labels the Cauchy surfaces $\Sigma_{\mathfrak{t}}$ used to foliate the Minkowski spacetime $(\mathcal{M}, g_{ab})$,
${\bf x}\equiv (x^1,x^2,x^3)$ are coordinates on $\Sigma_{\mathfrak{t}}$,
$\nabla_a$ is the torsion-free  covariant derivative compatible with the spacetime metric $g_{ab}$, $g \equiv {\rm det} (g_{\mu\nu})$ in some arbitrary coordinate system,  $\hat{\sigma}^{\rm z}$ is the  Pauli matrix associated with the detector internal states, $\hat{\sigma}^z|\pm\rangle=\pm|\pm\rangle,$ 
$c \in C^\infty_0(\mathbb{R})$ 
is the time-dependent coupling constant with compact support given by the finite proper time  $T$, and  $\psi(\mathfrak{t},{\bf x})$ is a smooth real function satisfying  $\psi|_{\Sigma_{\mathfrak{t}}} \in C^\infty_0(\Sigma_{\mathfrak{t}})$ 
for all $\mathfrak{t}$, which models the fact that the detector interacts with the field only in some vicinity of its worldline. 

The total Hamiltonian of the detector--field system is 
\begin{equation}
\hat{H}\equiv \hat{H}_\phi + \hat{H}_{\rm int},
\label{totalH}
\end{equation}
where the detector free Hamiltonian, ${\hat H}_0$, does not appear because
we have chosen (with no loss of generality) that $E_+=E_-$ is null, so that
$\hat{H}_0|\pm\rangle=E_\pm |\pm\rangle =0$.
Here,
\begin{equation}
\hat{H}_\phi (\mathfrak{t})
\equiv 
\int_{\Sigma_{\mathfrak{t}}} d^3{\bf x} 
\left( 
\hat{\pi}(\mathfrak{t},{\bf x})\dot{\hat{\phi}}(\mathfrak{t},{\bf x}) - \mathcal{L}[\hat{\phi},\nabla_a\hat{\phi}] 
\right)
\label{canonicalH}
\end{equation}
is the canonical Hamiltonian of the free scalar field, where 
$$ 
\mathcal{L} [ \hat{\phi},\nabla_a\hat{\phi} ]
\equiv
-\frac{1}{2}\sqrt{-g} \nabla_a\hat{\phi} \nabla^a\hat{\phi}
$$
is the Lagrangian density,  $\hat{\pi}\equiv \partial\mathcal{L}/\partial \dot{\hat{\phi}}$ is the momentum canonically conjugated to $\hat{\phi}$ with
$``\; \dot{}\;" \equiv \partial_{\mathfrak{t}}$, and $\hat{H}_{\rm int}$ is the interaction Hamiltonian which, in the interaction picture, is given by
\begin{equation}
\hat{H}_{\rm int}^I(\mathfrak{t}) \equiv c(\mathfrak{t})\int_{\Sigma_{\mathfrak{t}}}d^3{\bf x} \sqrt{-g} \psi(\mathfrak{t},{\bf x})\hat{\phi}(\mathfrak{t},{\bf x}) \otimes\hat{\sigma}^{\rm z}.
\label{Hint}
\end{equation} 

We can write the interaction picture time evolution operator with respect to the foliation $\Sigma_\mathfrak{t}$ as  \begin{equation}
\hat{U}=\mathcal{T}\exp{\left[-i\int_{-\infty}^\infty dt \hat{H}^I_{\rm int}(\mathfrak{t})\right]}, 
\label{unitaryevo}
\end{equation} 
where $\mathcal{T}$ indicates time ordering with respect to $\mathfrak{t}$. By making use of the Magnus expansion~\cite{magnus}
\begin{equation}
\hat{\Omega}\equiv\sum_{n=1}^\infty\hat{\Omega}_n,
\label{magnus1}
\end{equation}
where each $\hat{\Omega}_n$ is an operator of order $n$ in 
$\hat{H}^I_{\rm int}(\mathfrak{t})$ 
with 
\begin{eqnarray}
\hat{\Omega}_1
&=&
-i\int_{-\infty}^\infty d\mathfrak{t} \hat{H}^I_{\rm int}(\mathfrak{t}), \label{omega1}\\
\hat{\Omega}_2
&=&
-\frac{1}{2}
\int_{-\infty}^\infty d\mathfrak{t}
\int_{-\infty}^{\mathfrak{t}} 
d\mathfrak{t}'
\left[ 
\hat{H}^I_{\rm int}(\mathfrak{t}),\hat{H}^I_{\rm int}(\mathfrak{t}')
\right] 
\\
\hat{\Omega}_3
&=&
\frac{i}{6}
\int_{-\infty}^\infty \!\!\!\! d\mathfrak{t} \int_{-\infty}^{\mathfrak{t}}\!\!\!\! d\mathfrak{t}' \int_{-\infty}^{\mathfrak{t}'}\!\!\!\! d\mathfrak{t}'' \left(\left[\hat{H}^I_{\rm int}(\mathfrak{t}) ,\left[\hat{H}^I_{\rm int}(\mathfrak{t}'),\hat{H}^I_{\rm int}(\mathfrak{t}'')\right] \right] \right. 
\nonumber \\ 
&+& 
\left.
\left[
\hat{H}^I_{\rm int}(\mathfrak{t}'') ,
\left[
\hat{H}^I_{\rm int}(\mathfrak{t}'),
\hat{H}^I_{\rm int}(\mathfrak{t})
\right]
\right]
\right),
\label{omega3}
\end{eqnarray}
and with the high-order terms being obtained recursively, we can cast Eq.~(\ref{unitaryevo}) as
\begin{equation}
\hat{U}=\exp{\hat{\Omega}}. 
\label{magnus2}
\end{equation}
By using Eq.~(\ref{Hint}) and the covariant canonical commutation relation 
\begin{equation}
[\hat{\phi}(x),\hat{\phi}(x')]\equiv-i\Delta(x,x')\hat{I}
\label{CCR}
\end{equation}
in Eqs.~(\ref{omega1})-(\ref{omega3}), we can write
\begin{eqnarray}
\hat{\Omega}_1&=&-i\hat{\phi}(j)\otimes  \hat{\sigma}^{\rm z}, \label{Omega1}\\
\hat{\Omega}_2&=& i \; \Xi \hat{I},
\label{Omega2}\\
\hat{\Omega}_k &=& 0, \; \mathrm{for}\; k\geq 3. 
\label{Omegak}
\end{eqnarray}
Here,
\begin{equation}
\hat{\phi}(j)\equiv \int_{\mathcal{M}}d^4x\sqrt{-g}j(x)\hat{\phi}(x)
\label{smeared}
\end{equation}
with
\begin{equation}
j(\mathfrak{t},{\bf x})\equiv c(\mathfrak{t}) \psi(\mathfrak{t},{\bf x})
\label{f}
\end{equation}
is a compact-support function on $\mathcal{M}$ carrying the information about the detector, and
\begin{equation*}
\Xi 
\equiv 
\frac{1}{2}
\int_{-\infty}^\infty d\mathfrak{t} \; c(\mathfrak{t})
\int_{-\infty}^{\mathfrak{t}}  d\mathfrak{t}' c(\mathfrak{t}') \Delta(\mathfrak{t},\mathfrak{t}')
\end{equation*}
with 
$$
\!\Delta(\mathfrak{t},\mathfrak{t}')
\equiv\!\! 
\int_{\Sigma_{\mathfrak{t}}}\!\!\!\! d^3{\bf x}\sqrt{-g} \int_{\Sigma_{\mathfrak{t}'}}\!\!\!\! d^3{\bf x'} \sqrt{-g'}\psi(\mathfrak{t},{\bf x})\Delta(x,x')\psi(\mathfrak{t}',{\bf x'}).
$$ 
As a result, using Eqs.~(\ref{Omega1})-(\ref{Omegak}) in Eqs.~(\ref{magnus1}) and~(\ref{magnus2}),  the unitary evolution of the system becomes
\begin{equation}
\hat{U}=e^{i\Xi}e^{-i\hat{\phi}(j)\otimes \hat{\sigma}^{\rm z}}.
\label{unitaryfinal}
\end{equation}

To compute the emission rate of a pointlike uniformly accelerated gapless UD detector with respect to inertial observers, let us take~\cite{finalnote} 
\begin{equation}
j=\left\{\begin{array}{cc} c_0\delta(\xi)\delta^2({\bf x}_\bot) & -T/2< \tau <T/2  \\ 0 & |\tau|> T/2\end{array}\right. , \quad c_0={\rm const}
\label{source}
\end{equation}
where we recall that $(\tau, \xi, {\bf x}_\bot)$, with $\tau,\xi \in \mathbb{R}$ and ${\bf x}_\bot=(x,y)\in \mathbb{R}^2$, are Rindler coordinates covering the right Rindler wedge (region $z>|t|$).  

In addition, let us expand the field operator as (see, {\em e.g.,} Ref.~\cite{LFM19}) 
 \begin{eqnarray}
 \hat{\phi}(x)
 &=& 
 \sum_{\sigma=1}^{2} \int_0^\infty d\omega_R 
 \int d^2{\bf k}_\perp 
 \left[
 w^\sigma_{\omega_R {\bf k}_\perp}(x) \hat{a}(w^{\sigma *}_{\omega_R {\bf k}_\perp})\right. 
 \nonumber \\
 &+&  
 \left. w^{\sigma *}_{\omega_R {\bf k}_\perp}(x) \hat{a}^\dagger(w^\sigma_{\omega_R {\bf k}_\perp})
 \right],
 \label{unruhout}
 \end{eqnarray}
where 
$ \hat{a}(w^{\sigma *}_{\omega_R {\bf k}_\perp})$ 
and 
$\hat{a}^\dagger(w^\sigma_{\omega_R {\bf k}_\perp})$, $\sigma=1,2,$ are the annihilation and creation operators of the so called Unruh modes 
$$
\left\{w^1_{\omega_R {\bf k}_\bot}, 
w^2_{\omega_R {\bf k}_\bot} \right\},
$$ 
where
\begin{eqnarray}
w^1_{\omega_R {\bf k}_\perp} 
& \equiv &
\frac{u_{\omega_R {\bf k}_\perp}^R + 
e^{-\pi\omega_R/a}u_{\omega_R\,-{\bf k}_\perp}^{L*}}
{\sqrt{1-e^{-2\pi\omega_R/a}}},
\label{wpositive1}\\
w^2_{\omega_R {\bf k}_\perp} 
& \equiv &
\frac{u_{\omega_R {\bf k}_\perp}^L + e^{-\pi\omega_R/a}u_{\omega_R\, 
-{\bf k}_\perp}^{R*}}{\sqrt{1-e^{-2\pi\omega_R/a}}}.
\label{wpositive2}
\end{eqnarray} 
Such modes (together with their Hermitian conjugates) form a complete orthonormal set of solutions of the Klein-Gordon equation and are positive-frequency with respect to the inertial time $t$. 
We recall that  $u_{\omega_R {\bf k}_\perp}^R $ are the right-wedge Rindler modes~(\ref{REF}) vanishing in the left wedge, whereas  
$u_{\omega_R {\bf k}_\perp}^L\equiv u_{\omega_R {\bf k}_\perp}^{R*} (-t,x,y,-z)$ are the corresponding left-wedge Rindler modes. 

The fact that modes~(\ref{wpositive1}) and~(\ref{wpositive2}) are positive-frequency with respect to the inertial time but are labeled by Rindler quantum numbers, $\omega_R$ and ${\bf k}_\perp$, makes them particularly suitable to scrutinize the relation between the radiation seen by inertial observers and the physics developed by uniformly accelerated observers.

Now, by smearing Eq.~(\ref{unruhout}) with $j$, one obtains~\cite{LFM19} 
 \begin{equation}
 i\hat{\phi}(j)=  \hat{a}^\dagger\left(KEj\right) -\hat{a}\left({KEj}^* \right),
 \label{unruhoutsmeared2}
\end{equation}
where
\begin{equation}
\hat{a}^\dagger\left(KEj\right)
\!=\!\sum_{\sigma=1}^2 \int_0^\infty \!\!\!d\omega_R \!\int \!\!d{\bf k}_\perp  
\langle w^{\sigma}_{\omega_R {\bf k}_\perp}, Ej\rangle_{{}_{\rm KG}}\hat{a}^\dagger (w^{\sigma}_{\omega_R {\bf k}_\perp})
\label{adagger}
\end{equation}
and $Ej\equiv Aj -Rj$, with $Aj$ and $Rj$ being the advanced and retarded solutions of the Klein-Gordon equation with source $j$, respectively. 
Equation~(\ref{unruhoutsmeared2}) will allow us to evolve the system through Eq.~(\ref{unitaryfinal}). 

Now, it is convenient to use
\begin{equation}
w^1_{-\omega_R {\bf k}_\perp}=w^2_{\omega_R {\bf k}_\perp}
\label{w1w2}
\end{equation}
and take the limit $T\rightarrow \infty$ to cast Eq.~(\ref{adagger}) in the form~\cite{LFM19}
\begin{equation}
\!\hat{a}^\dagger\left(KEj\right)
=\frac{ic_0}{\sqrt{2\pi^2a}}\int d^2{\bf k}_\bot K_0(k_\bot/a)\hat{a}^\dagger\left(w^2_{0{\bf k}_\bot}\right),
\label{quantumKEj}
\end{equation}
where $a$ equals the detector proper acceleration. 

Now, let us assume that  in the asymptotic past  the UD detector is prepared in the state described by the density matrix $\hat{\rho}^D_{-\infty}$, while the field is in the Minkowski vacuum $|0_M\rangle$. 

The final state obtained in the asymptotic future after time-evolving the system with Eq.~(\ref{unitaryfinal}) will be given by
\begin{equation}
\hat{\rho}^{\phi D}_\infty \equiv \hat{U} |0_M\rangle \langle 0_M|\otimes \hat{\rho}^D_{-\infty} \hat{U}^\dagger.
\label{finalstate}
\end{equation}
As any mixed state can be written as a statistical mixture
\begin{equation}
\hat{\rho}^D_{-\infty}=\sum_{m=1}^{\mathfrak{D}} p_m|\varphi_m\rangle \langle \varphi_m|,
\label{qubitstatepast}
\end{equation} 
with  $\mathfrak{D}\in\mathbb{N}$, $p_m\in\mathbb{R}_+$, and $\sum_mp_m=1$, in order to compute the final state~(\ref{finalstate}) it will be enough to calculate
\begin{equation}
|\Psi^m_\infty\rangle \equiv \hat{U}|0_M\rangle\otimes|\varphi_m\rangle, 
\label{psiinfty}
\end{equation}
where $|\varphi_m\rangle \equiv \alpha_m |+\rangle + \beta_m |-\rangle$ and $|\alpha_m|^2+|\beta_m| ^2=1.$ Therefore, by using Eqs.~(\ref{unruhoutsmeared2}) and~(\ref{quantumKEj}) in Eq.~(\ref{unitaryfinal}) we can cast Eq.~(\ref{psiinfty}) as
\begin{equation}
|\Psi^m_\infty\rangle = \alpha_m |0(j)\rangle\otimes|+\rangle + \beta_m |0(-j)\rangle\otimes|-\rangle,
\label{final_state}
\end{equation} 
with
\begin{eqnarray}
&&|0(\pm j)\rangle
\equiv 
\exp [-T c_0^2a/4\pi^2] 
\nonumber \\
&&\times 
\exp 
\left[ \mp i\int d^2{\bf k}_\bot  
\frac{ c_0K_0(k_\bot/a)}{\sqrt{2\pi^2a}}\hat{a}^\dagger\left(w^2_{0{\bf k}_\bot}\right) 
\right] |0_M\rangle, \nonumber \\
\label{in-out3}
\end{eqnarray}
where $T$ should be understood as in Eq.~(\ref{T}), meaning the limit of a finite-time detector with $T$ made arbitrarily large.  It is important to note that Eq.~(\ref{in-out3}) is a (multi-mode) coherent state built entirely from zero-energy Unruh modes whose field expectation value in the asymptotic future is 
\begin{equation}
\langle 0(\pm j)|\hat{\phi}(x)|0(\pm j)\rangle=\pm Rj,
\label{FEVj}
\end{equation}
where we recall that $\pm Rj$ is precisely the classical retarded solution associated with the scalar source $\pm j$~\cite{LFM19}. 

Next, by using Eqs.~(\ref{qubitstatepast})-(\ref{in-out3}) in Eq.~(\ref{finalstate}) and tracing out the detector's degrees of freedom, one obtains for the field state in the asymptotic future
\begin{eqnarray}
\hat{\rho}^\phi_{\infty}
&=&
{\rm tr}_D [\hat{\rho}^{\phi D}_{\infty} ]
\nonumber \\
&=&
\sum_{m=1}^{\mathfrak{D}} 
p_m |  \psi^m_\infty \rangle \langle \psi^m_\infty |
\nonumber \\
&=&
\sum_{m=1}^{\mathfrak{D}} 
p_m\left(|\alpha_m|^2|0( j) 
\rangle \langle 0(j)| 
\right. 
\nonumber \\
&+&
\left. |\beta_m|^2|0(-j)\rangle \langle 0(-j)|\right), 
\label{fieldstatefinal}
\end{eqnarray}
which is a statistical mixture of $|0( j)\rangle $ and $|0(-j)\rangle$. 

The detector emission rate is, then, 
\begin{equation}
\Gamma^{\rm total}_{\rm em}\equiv {\rm tr} \left(\hat{\rho}^\phi_{\infty} \hat{N}\right)/T,
\label{ratemixed}
\end{equation}
where 
\begin{equation}
\hat{N}\equiv \int_{-\infty}^\infty d\omega_R \int d {\bf k}_\perp \hat{a}^\dagger (w^{2}_{\omega_R {\bf k}_\perp})\hat{a}(w^{2 *}_{\omega_R {\bf k}_\perp})
\label{Ntot}
\end{equation}
is the total (inertial-frame) particle number operator and we have used
Eq.~(\ref{w1w2}) in order to cast $\hat{N}$ in the form above. In order to compute it, we first use Eqs.~(\ref{in-out3}) and~(\ref{Ntot}) to get 
\begin{equation}
\frac{\langle 0(\pm j)| \hat{N}| 0(\pm j)\rangle}{T}= \frac{c_0^2 a}{4\pi^2}
 \label{NMrate3}
 \end{equation}
and then use Eqs.~(\ref{fieldstatefinal}) and~(\ref{NMrate3}) in Eq.~(\ref{ratemixed}) to recover the emission rate for a classical (scalar) charge with proper acceleration $a$, Eq.~(\ref{totalMinkowskirate}):
\begin{eqnarray}
\Gamma^{\rm total}_{\rm em} 
&\equiv&
\sum_{m=1}^{\mathfrak{D}} p_m|\alpha_m|^2 
\frac{\langle 0( j)| \hat{N}| 0(j)\rangle}{T} 
\nonumber \\
&+& 
\sum_{m=1}^{\mathfrak{D}} p_m|\beta_m|^2
\frac{\langle 0(-j)| \hat{N}| 0(-j)\rangle}{T} 
\nonumber \\
&=&  \frac{c_0^2 a}{4\pi^2},
 \label{NMrate4}
 \end{eqnarray}

Moreover, not only a uniformly accelerated gapless detector emits particles like a scalar source, but the energy content of such particles is also indistinguishable from that of a uniformly accelerated source. This can be seen by noting that the (normal-ordered) stress-energy tensor: 
  \begin{eqnarray}
: \hat{T}_{ab}: \; \equiv \hat{T}_{ab}-\langle 0_M|  \hat{T}_{ab} | 0_M\rangle,
 \label{normalTab}
 \end{eqnarray}
 with
 $
 \hat{T}_{ab}
 =\nabla_a\hat{\phi}\nabla_b\hat{\phi}-\frac{1}{2}g_{ab}\nabla^c\hat{\phi}\nabla_c\hat{\phi},
 $
 satisfies 
 \begin{eqnarray}
\!\!\!\!\!\!\!\!\!\!\!\!\!\!\!\!
\langle 0(\pm j)|: \hat{T}_{ab}:| 0(\pm j)\rangle
\!\!\! &=& \!\!\!
\nabla_aR(\pm j)  \nabla_bR(\pm j) 
\nonumber \\
&-& 
\frac{g_{ab}}{2}\nabla^c R(\pm j) \nabla_c R(\pm j). 
\label{<normalTab>}
\end{eqnarray}
As a result, using Eqs.~(\ref{fieldstatefinal}) and~(\ref{<normalTab>}), we have
\begin{eqnarray}
\!\!\!\!\!\!\!\!\langle : \hat{T}_{ab}:\rangle_{\rho^\phi_\infty}
&\equiv& 
{\rm tr}\left(\rho^\phi_\infty : \hat{T}_{ab}: \right)
\nonumber \\
 &=& 
\nabla_aRj  \nabla_bRj -\frac{1}{2}g_{ab}\nabla^c Rj \nabla_c Rj. 
 \label{<normalTab>mixed}
 \end{eqnarray}
which is exactly the {\rm classical} stress-energy tensor, $T_{ab}[Rj],$ associated with the retarded solution $Rj$. Hence, if one computes, {\em e.g.,}  the energy flux integrated along a large sphere in the asymptotic future we obtain~\cite{RW94}
\begin{equation}
\int d S^b \langle : \hat{T}_{ab}:\rangle_{\rho^\phi_\infty}  (\partial_t)^a = \frac{c_0^2a^2}{12\pi},
\label{scalarlarmor}
\end{equation}
which agrees with the Larmor formula for the power radiated by a scalar source (with respect to inertial observers). Here, $d S^b$ is the vector-valued volume element on the sphere and $(\partial_t)^a$ is the Killing field associated with a global inertial congruence. 

\section{A close look at mode functions of small or zero frequency}
\label{app:zeromode}

	The  function (\ref{RM})   consists of three  factors:
\begin{eqnarray}
 A(\omega) &= \sqrt{\sinh(\pi \omega/a)/4 \pi^4 a},\\
B(\omega,k_\bot,\xi) &= K_{i\omega/a}(k_\bot e^{a\xi}/a),
\end{eqnarray}
and a plane wave.
$A$ is  a normalization factor chosen to trivialize the integration density in the eigenfunction expansion --- in other words, to make the inner product of two of these modes equal to exactly $\delta(w - w')$ \cite[p.~796]{CHM08}.  We note that $A \to  0$ as $\omega\to 0$.

The differential equation satisfied by $B$ \cite[(2.86)]{CHM08} has a ``classical turning point''  where $k_\bot^2 e^{2a\xi} = \omega^2$.  If $k_\bot \ne 0$, the solution satisfies $B''/B < 0$ if $\xi$ is to the left of that point, and $B''/B > 0$ to the right, where $``\;'\; "\equiv d/d\xi$.  Thus, $B$ changes from oscillatory to decaying at that point,  $a\xi = \log(\omega/k_\bot)$  (assuming $\omega$ and $k_\bot$ positive).  As $w \to 0$,  $\xi \to -\infty$.  

Furthermore, $K_\nu(z) = \sqrt{\pi/(2z)}\, e^{-z} + O(z^{-2}e^{-z})$  
 \cite[(8.451.6)]{GR80}.    Thus the leading term in $B$ is independent of $\omega$  (though 
 the higher-order terms grow with $\omega$).

 As $w \to  0$ with fixed $k_\bot \ne 0$, 
 three things happen.  Of course, the time dependence of (\ref{RM}) becomes very slow, and in the  limit the function is static. But also, 
 the mode function becomes ``small'' in the Rindler region R in two 
different senses.  
 
 First, any fixed $\xi \ne 0$ eventually falls in the regime of 
exponential decay, and $B$ is increasingly concentrated in the region of smaller (i.e., more negative) $\xi$.    When $\omega = 0$, 
 $B(0,k_\bot,\xi) = K_0(k_\bot e^{a\xi}/a)$ .  The 
$K_0$ function is decaying everywhere, starting from a logarithmic 
singularity at $e^\xi = 0$.  
The corresponding full mode function is independent of (Rindler) time, 
and it does satisfy the wave equation.  

Second, the normalized full mode function vanishes as $\omega\to 
0$ because of the $A$ factor.  When it appears (at an endpoint) 
in an integral over $\omega$, however, the other factors in the 
integrand may be singular as $\omega \to 0$ in such a way that 
the interval around $\omega = 0$ makes a nontrivial contribution 
to the integral.  This is what happens in the calculations with 
the Unruh bath in Sec.\ \ref{ssec:IIB}.
  Even in the limit $\Delta E \to 0$, the nonvanishing function $B(0,k_\bot ,\xi)$ makes its presence felt.
  
  The vanishing of $A(0)$ should not be interpreted as saying that ``no mode with $\omega=0$ exists.''
  Just like any other point in a continuous spectrum, $\omega=0$ may indeed be omitted
  from any spectral integral without changing the result.  But the contribution from any
  small interval, $[\omega, \omega+\epsilon]$,  of continuous spectrum is significant,
  no matter whether $\omega=0$ or $\omega\neq 0$.
  By partial analogy, $\omega=0$ is not missing from a Fourier sine transform
  $$f(x) = \frac2{\pi}\int_0^\infty \hat{f}_S(\omega) \sin(\omega x)$$
just because $\sin(0x)=0$.



\begin{thebibliography}{}
\bibitem{polar} 
    The situation is analogous to the normal modes of the wave equation in Minkowski space in spherical coordinates, 
    $\sqrt{\omega/\pi}\, j_l(\omega r) Y_{lm} (\theta, \phi) e^{-i\omega t}$, which exist for arbitrarily large $l$ and arbitrarily small $\omega$, even reducing to $\omega^{1/2+l} r^l Y_{lm}(\theta, \phi) e^{-i\omega t}/(2^{l+1} \Gamma [3/2+l])$ for $\omega \approx 0$. In that case, the modes with small $\omega$ and large $l$ concentrate at infinity rather than at an horizon.
\bibitem{note}
    These ``proper conditions''  are  mostly related to the fact that the charge must have enough time to thermalize in the Unruh thermal bath. Although the experiment is feasible under present technology, it has not been realized yet to the best of our knowledge.
\bibitem {finalnote} 
Here, it is implicitly assumed that the source is switched on/off continuously. However, because we will take the limit $T\rightarrow \infty$ before radiation rates are evaluated, we will be able to avoid any divergences without worrying about how the detector is switched on/off.
\end{thebibliography}
    \end{document}